\documentclass[twoside]{LCWS11}
\usepackage{slashed} 
\usepackage{nccmath} 
\usepackage{subfigure} 
\usepackage[latin1]{inputenc}
\usepackage[dvips]{graphicx,epsfig,color}
\usepackage{wrapfig,rotating}
\usepackage{amssymb,amsmath,array}
\usepackage{cite}
\newcommand{\lp}{\left(} \newcommand{\rp}{\right)} 
  
\newcommand{\ls}{\left[}  \newcommand{\rs}{\right]}

\DeclareMathOperator{\Ai}{Ai}

\DeclareMathOperator{\e}{e}

\begin{document}
\title{A closer look at the beam-beam processes at ILC and CLIC}
\author{A Hartin$^1$
\vspace{.3cm}\\
1- DESY, Notkestrasse 85, 22607 Hamburg, Germany}


\maketitle

\begin{abstract}The strength of the electromagnetic fields in the bunch collision at a linear collider will have a significant effect, yielding large numbers of beamstrahlung photons and associated coherent pair production. These effects are limited in the proposed ILC beam parameters which limit the strength of the bunch field to $\Upsilon_{\text{ave}}=0.27$. The CLIC 3 Tev design by comparison has a $\Upsilon_{\text{ave}}=3.34$ yielding huge number of coherent pairs. In terms of the precision physics programs of these proposed colliders there is an imperative to investigate the effect of the strong bunch fields on higher order processes. From the exact wavefunctions used in the calculation of transition rates within the Furry interaction picture, and using appropriate simplifications, a multiplicative factor to the coupling constants was obtained. This indicates a significant variation to the transition rate near threshold energies. Further studies are in progress to calculate the exact effect on expected observables.
\end{abstract} 

\section{Introduction}

Within the beam-beam collision at the Interaction Point of a collider, strong electromagnetic fields are present which can affect particle processes. The parameter that is used to estimate the impact of the strong fields on particle transition rates is the dimensionless $\Upsilon$ parameter - the ratio of the field strength to the Schwinger critical field at which real pairs are produced from the vacuum. Beam parameters are chosen in the ILC design studies which limit $\Upsilon$ to a value below unity which inhibits the production of coherent background pairs. The CLIC design parameters at 3 TeV center of mass energy allow $\Upsilon$ to range up to the value of 10, producing a large number of coherent pairs per bunch crossing at the level of a significant fraction of the original bunch population. \\

Coherent pair production at a collider is a one vertex process which is kinematically allowed because of the contribution of momentum from the external field. The transition rate is calculated in the Furry interaction picture which utilises special wavefunctions for particles that couple to the external field. The dependence on the external field strength is a reasonably complex integration over special functions and polynomials. In order to gauge the effect on higher order processes it is necessary to carry out explicit calculations within the Furry picture. \\

The calculation of two-vertex processes in an external field is an ongoing field of study. In a circularly polarised field, among other studies, Moller scattering \cite{Oleinik67,Oleinik68,Bos79a,Bos79b}, Compton Scattering and Pair Production \cite{Hartin06} have been considered and cross-sections reveal themselves as resonant, exceeding one vertex cross-sections at the resonances. \\

At a collider, the external field produced by a relativistic charge bunch is a constant crossed electromagnetic field. A consideration of two vertex Furry picture processes in such a field is required in order to gauge the effect on collider process cross-sections. Furry picture helicity amplitudes will be particularly important to understand precision spin physics such as measurement of the triple gauge coupling and in the use of the W pair production to aid in the calculation of the luminosity weighted polarisation as accurately as possible. \\

In this paper some of the building blocks of this strong field physics program will be assembled. Simulations of the beam-beam interaction at ILC and CLIC reveal the numerical value of the $\Upsilon$ parameter and the luminosity weighted depolarisation due to lowest order processes. The wavefunctions required for calculation of the W pair production in an external field are introduced and an estimate of the dependence of the cross-section on the $\Upsilon$ parameter is obtained. 

\section{1st order coherent processes at a collider}

Each incoming charged particle to the interaction point of a collider sees the electromagnetic field of the oncoming bunch. It is possible to completely take into account the strong bunch field by solving the Dirac equation containing the field potential \cite{Volkov35}. With use of these solutions in the Feynman diagram formalism, the transition rates of the first order processes - the beamstrahlung and the coherent pair production - can be calculated. \\

When helicity amplitudes are used, the spin dependent beamstrahlung transition rate for an electron of spin $s_\nu$ momentum $p_\mu$ in a field with field tensor $F^{\mu\nu}$, momentum $k_\mu$ and potential $a_\mu$ can be written \cite{Ritus72},

\begin{gather}\label{beamstr}
 W= \mfrac{\alpha m^2}{\pi\epsilon_p}\int_0^{\infty}\mfrac{du}{(1+u)^2}\ls \mfrac{e}{m^3}F^{*\mu\nu}p_\mu s_\nu \mfrac{z\Ai(z)}{1+u}-\Ai_1(z)-\mfrac{2+2u+u^2}{z(1+u)}\Ai'(z)\rs \\
 \text{where}\quad z=\lp\mfrac{u}{\Upsilon}\rp^{2/3}, \quad \Upsilon=\mfrac{e|\vec{a}|(k\cdot p)}{m^3} \notag
\end{gather}

Equation \ref{beamstr} gives the spin flip rate which, along with spin precession, leads to IP depolarisation of the electron and positron beams. The pertinent ILC and CLIC parameters along with the expected depolarisation are shown in table \ref{tab1}. \\

\begin{table}
\centerline{\begin{tabular}{|c|| c | c |}\hline
 Parameter & ILC 1TeV & CLIC 3 TeV \\ \hline\hline
    $\mathcal{L} (\times10^{34})$ & 4 & 3.6 \\ \hline
   N(incoh) & 3.9e5 & 3.8e5 \\ \hline
   N(coh) & 0 & 6.8e8  \\ \hline
   $\Upsilon$ (ave) & 0.27 & 3.34 \\ \hline
   $\Upsilon$ (max) & 0.94 & 10.9 \\ \hline
   $\delta E_{\text{bs}}$ & 10\% & 28\% \\ \hline
   $\langle\text{depol}\rangle_{\text{LW}}$& 0.62\% & 3.5\% \\ \hline
\end{tabular}}\caption{\bf ILC and CLIC IP beam parameters}\label{tab1}\end{table}

\begin{figure}[h] 
\centerline{\includegraphics[width=0.45\textwidth,angle=270]{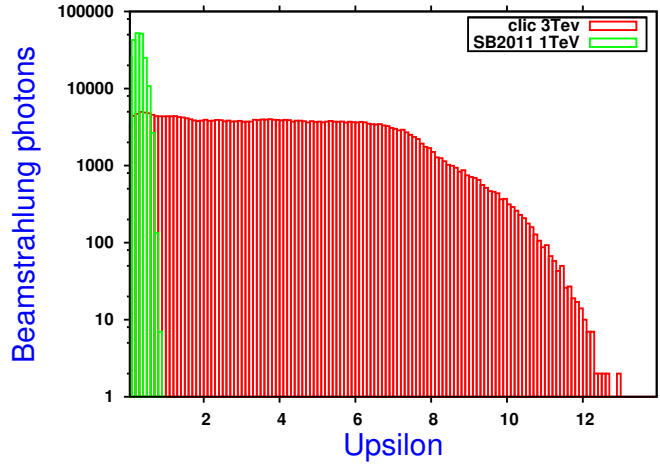}}
\caption{\bf Upsilon value of coherent pairs.}\label{upsvals}
\end{figure}

The $\Upsilon$ value varies during the IP bunch collision as the bunches distort under the pinch effect. Using a tracking program such as CAIN \cite{Yokoya03}, $\Upsilon$ for each photon involved in a coherent pair production can be determined and a distribution plotted (figure \ref{upsvals}). This distribution sets the scale of $\Upsilon$ values to take into account.

Since CLIC produces copious numbers of coherent pairs and ILC very little, the coherent pair production transition rate which is related to that of equation \ref{beamstr} is clearly limited by the numerical value of $\Upsilon$. One can check in fact that as $\;\Upsilon\!\rightarrow\!0\;$ the transition rate vanishes as expected. \\

A non-zero $\Upsilon$ will have an effect on all physics processes at a collider and it is important to determine that effect by a consideration of higher order external field processes.

\section{Exact wavefunctions in the external field}

In the Furry picture \cite{Furry51} (see also the Appendix of \cite{hartin11}) one separates off the external field potential $A^e$ from the part of the Lagrangian containing the electromagnetic interactions and situates it in the equations of motion of the various charged particles. In particular, for fermion wavefunctions $\psi^e_p$ with rest mass $m_e$ and W bosons wavefunctions $W^{e\pm}$ with rest mass $m_W$ the equations of motion are \cite{Kurilin99} ($D^\pm=\partial_\mu\pm ieA^e_\mu$),

\begin{align}\label{eq:dirac}
\lp D^{\pm2}+m_e^2 \pm \mfrac{e}{2}\sigma^{\mu\nu}F_{\mu\nu} \rp \psi_p^e=0 \notag\\[6pt]
\lp D^{\pm2}+m_W^2 \rp W^{e\pm\mu} \pm 2ieF^{\mu\nu} W^{e\pm}_\nu=0
\end{align}

Exact solutions of these equations of motion are possible for the case of plane wave electromagnetic fields such as those present in the beam-beam interaction. The resulting wavefunctions differ from the usual solutions by an extra phase $S_p$ and modified spinors $E_p, E_W$,

\begin{gather}\label{eq:Volkov}
 \psi_p^e = \,\mfrac{1}{\sqrt{2\epsilon_pV}}\,E_\psi\, e^{-i\lp p\cdot x+S_p(k\cdot x)\rp} \notag\\
 W_p^e = \,\mfrac{1}{\sqrt{2\epsilon_pV}}\,E_W\, e^{-i\lp p\cdot x+S_p(k\cdot x)\rp} \\[4pt]
\text{where}\hspace{0.5cm} E_\psi = \ls 1-\mfrac{e\slashed{A}^e\slashed{k}}{2(k\cdot p)}\rs u_p \; ,
\hspace{0.5cm} E_W = \ls g_{\mu\nu} \pm \mfrac{e}{(k\cdot p)}F_{\mu\nu}-\mfrac{e\slashed{A}^e\slashed{k}}{2(k\cdot p)^2} k_\mu k_\nu \rs v^{\pm\nu}_p  \notag \\[4pt] 
S_p(k\cdot x)\equiv\int_{0}^{(k\cdot x)}\ls \pm\mfrac{e(A^{e}(\phi)\cdot p)}{(k\cdot p)}-\mfrac{e^2A^e(\phi)^2}{2(k\cdot p)}\rs d\phi \notag
\end{gather}

\section{The Furry picture vertex}

\begin{figure}[h]
\centerline{\includegraphics[width=0.35\textwidth]{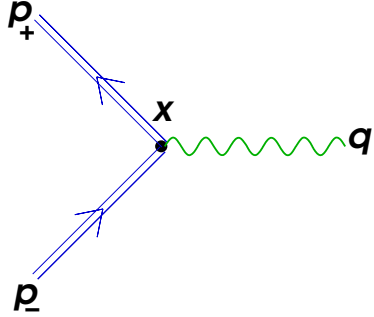}}
\caption{\bf The external field QED vertex.}\label{schanvert}
\end{figure}

In order to proceed to a transition rate, one plugs in the exact wavefunctions into the external field (equation \ref{eq:Volkov}) with the aid of Feynman diagrams in the usual way. For two vertex processes, such a calculation is quite complex and and its details require a separate study. Instead, here we examine the initial vertex of an s-channel process (figure \ref{schanvert}) in order to estimate the dependence on the strong field $\Upsilon$ parameter

To simplify the calculation (since we are only making an estimate here) we assume that an incoming electron and positron in the centre of mass frame $\sqrt{s}=2 E$ each see an anti-collinear field of equal strength so that the scalar products,

\begin{gather}
(A\cdot p)=0, \quad (k\cdot p)=2 E 
\end{gather}

Substituting into the wavefunction solutions the explicit expression for the constant crossed field 4-potential, $A^e_\mu=a_\mu (k.\cdot x)$ and transforming to momentum space, the Furry picture fermion wavefunction becomes,

\begin{gather}\label{psimomspace}
 \psi_p^e = \,\mfrac{1}{\sqrt{2\epsilon_pV}} 
\int dr\e^{-i(p+r\xi m_e\hat{k})\cdot x} \ls 1+i\mfrac{\chi}{\xi}\slashed{\hat{a}}\,\slashed{\hat{k}}\mfrac{d}{dr}\rs \Ai(r) \; u_p \notag \\[4pt]
\text{where}\quad \xi=-\mfrac{m_e}{2^{4/3}E}\Upsilon^{2/3}, \quad \chi=\mfrac{m_e^2}{8E^2}\Upsilon 
\end{gather}

Now considering that for the ILC and CLIC design parameters, $\;\Upsilon\approx 0.1\text{-}10\;$ and $\;E\approx 1\text{-}6\times 10^6\;$, the term in $\chi/\xi$ in equation \ref{psimomspace} can be neglected. A lower bound on the $r$ integration can be determined by requiring the energy of the bound state to be at least the fermion rest mass,

\begin{gather}\label{rlowbound}
r \geq \mfrac{1-\gamma}{\xi}, \quad \gamma=E/m_e
\end{gather}

Since there are two fermion wavefunctions contributing to the vertex of figure \ref{schanvert},  something like the square of an integration over $r$ can be expected. A multiplicative factor can be extracted from the Furry picture vertex which gives the numerical scale of the variation from the coupling constant in the normal interaction picture,

\begin{figure}
\centerline{\includegraphics[width=0.65\textwidth]{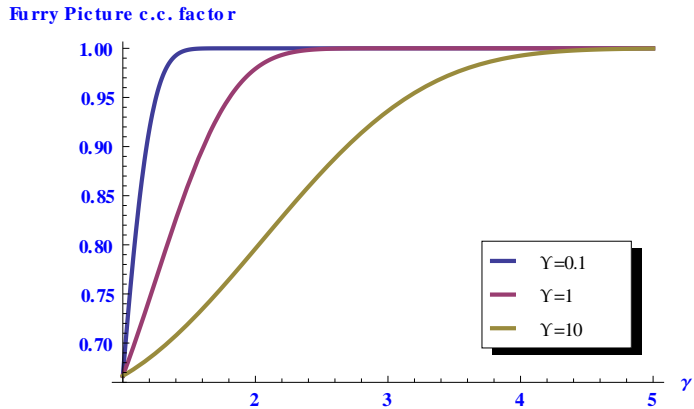}}
\caption{\bf The Furry picture coupling constant multiplicative factor.}\label{furpic_cc}
\end{figure} 

\begin{gather}\label{furcoupconst}
ig^e_e\gamma_\mu \equiv \ls \int^\infty_{\frac{1-\gamma}{\xi}}dr\Ai(r) \rs^2 ig_e\gamma_\mu
\end{gather}

As the external field vanishes, the lower bound of the integration in equation \ref{furcoupconst} approaches $-\infty$ and the integration itself, unity. That is, in the correct limit the Furry picture coupling constant approaches the normal coupling constant. \\

The multiplicative factor in equation \ref{furcoupconst} is plotted In figure \ref{furpic_cc} for various values of the $\Upsilon$ parameter expected for ILC and CLIC, and for the beam energy expressed by $\gamma$. The variation from the normal coupling constant is greatest near the threshold energy and for larger $\Upsilon$ values.
\section{Summary and Future work}
As a preliminary to further work, we have here gauged the effect of the strong bunch fields on physics processes at the ILC and CLIC. Simulations show that the $\Upsilon$ parameter which quantifies the field strength seen by individual particles, varies up to the value of 1 for ILC 1 TeV and 10 for CLIC 3 TeV beam parameters. By approximating the IP bunch collision as headon, and by considering only the leading term in the wavefunction solutions in the external field, a multiplicative factor to the normal coupling constants can be obtained. This multiplicative factor - in effect a Furry picture coupling constant - varies with particle energy and with $\Upsilon$ and is most significant around threshold energies. This is expected to be noticeable enough to warrant a complete study of particular processes that are sensitive to such changes. \\
 
Such a study would include different vertex factors for different species of particles. For instance the vertices involving W bosons will not show such a strong variation with field strength since both $\Upsilon$ and the the relativistic $\gamma$ are smaller for higher rest mass particles. Nevertheless, a transition product of a 2 vertex process in such a field is a product of all vertex factors, including the significantly affected $e^{+}e^{-}$ vertex, therefore all processes are expected to be affected. \\

A full transition rate calculation will allow for additional terms in the modified spinors, and the integration factor $r$ will also appear in Mandelstam variables and within the trace part. Moreover the particles involved in the process face not one but two external fields, from both of the incoming charge bunches. Though exact solutions in such combination of fields exist in only restricted configurations, progress is expected to be made by using the principle of superposition to simplify the problem to be solved. \\

Finally, real events will have to be generated for further analyses, such as detector analyses. Such an event generator will have to include a full electromagnetic tracker of the bunch collision in order to determine the $\Upsilon$ parameter at each point of production/decay. All this work is to appear in forthcoming studies.

\bibliographystyle{unsrt}
\bibliography{/home/hartin/Physics_Research/mypapers/hartin_bibliography}

\end{document}